\documentclass[%
 reprint,
showpacs,
 amsmath,amssymb,
 aps,
prb,
]{revtex4-1}

\usepackage{dcolumn}
\usepackage{bm}

\usepackage[dvipdfmx]{graphicx,color}

\usepackage{ulem} 

\def\<{\langle}
\def\>{\rangle}
\def\({\left(}
\def\){\right)}
\def\[{\left[}
\def\]{\right]}
\def\up{\uparrow}
\def\dn{\downarrow}
\def\s{\sigma}

\def\i{\mathrm{i}}

\newcommand{\cred}{\color{red}}

\newcommand{\abs}[1]{
\left\lvert {#1} \right\rvert}%
\newcommand{\paren}[1]{%
\left( {#1} \right)}
\newcommand{\sbra}[1]{
\left[ {#1} \right]}

\newcommand{\ds}[1]{
{\displaystyle {#1}}}

\begin{document}


\title{Origin of in-plane anisotropic resistivity in the antiferromagnetic phase of Fe$_{1+x}$Te} 

\author{Eiji Kaneshita}
\affiliation{National Institute of Technology, Sendai College, Sendai 989-3128, Japan}
\email{eiji@sendai-nct.ac.jp}
\author{Takami Tohyama}
\affiliation{Department of Applied Physics, Tokyo University of Science, Tokyo 125-8585, Japan}
\email{tohyama@rs.tus.ac.jp}

\date{\today}

\begin{abstract}
Motivated by a recent experimental report on in-plane anisotropic resistivity in the double-striped antiferromagnetic phase of FeTe, we theoretically calculate in-plane resistivity by applying a memory function approach to the ordered phase. We find that the resistivity is larger along an antiferromagnetically ordered direction than along a ferromagnetically ordered one, consistent with experimental observation. The anisotropic results are mainly contributed from Drude weight, whose behavior is attributed to Fermi surface topology of the ordered phase.
\end{abstract}



\maketitle



\section{Introduction}

Electronic states of two-dimensional Fe plane with square lattice are crucial for the mechanism of superconductivity in iron-based superconductors. At high temperature, the electronic state of the plane is isotropic without directional difference between two nearest-neighbor Fe-Fe directions of the square lattice. With decreasing temperature, an anisotropic electronic state emerges through the breakdown of four-fold symmetry in magnetic~\cite{Kasahara12Nature}, electric~\cite{Chu10Science,Tanatar10PRB, Ishida11PRB, Kuo11PRB, Ying11PRL, Ishida13PRL}, and electronic~\cite{Yi11PNAS,Dusza11EPL, Nakajima11PNAS, Nakajima12PRL, Chuang10Science, Allan13NP, Fu12PRL} properties, resulting in a nematic state with two-fold symmetry distinguishing two Fe-Fe directions. When antiferromagnetism next to superconductivity in phase diagrams appears, the anisotropy is strongly enhanced, for example, in resistivity measurements for detwinned samples of Ba(Fe,$TM$)$_2$As$_2$ with $TM=$Co~\cite{Chu10Science, Ishida13PRL}, Cr, and Mn~\cite{Kobayashi15JPSJ} as well as (Ba,K)Fe$_2$As$_2$~\cite{Ishida13JACS,Blomberg13NC}. Not only BaFe$_2$As$_2$ systems (called 122 system), but also Se- and Cu-substituted Fe$_{1+x}$Te systems (called 11 systems) exhibit in-plane anisotropy of resistivity~\cite{Liu15PRB}.

\begin{figure}[t]
\begin{center}
\includegraphics[width = 0.7\linewidth]{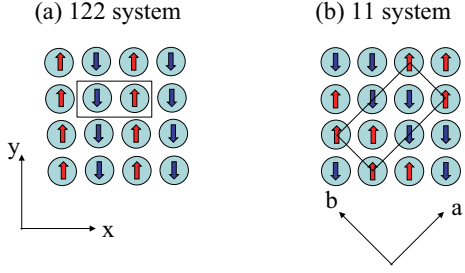}
\caption{(Color online) Schematic illustration of spin configurations for (a) 122 and (b) 11 systems. Axes $x$ and $a$ ($y$ and $b$) show the direction across (along) the stripes. The rectangles display the unit cell.
}
\label{fig:spin-config}
\end{center}
\end{figure}

In the 122 systems, antiferromagnetic (AFM) order occurs with a stripe-type spin arrangement characterized by ordering vector $\mathbf{Q}=(\pi,0)$, by defining the $x$ and $y$ directions to be nearest-neighbor Fe-Fe directions: AFM arrangement along $x$, ferromagnetic (FM) along $y$ [see Fig.~\ref{fig:spin-config}(a)].
The asymmetry gives rise to preference in electronic transport for the $x$ or $y$ direction.
It is very intuitively supposed that carriers will be scattered more strongly along the AFM-ordered direction than along the FM-ordered direction. This is tempting us to expect larger resistivity along the $x$ direction as compared with the $y$ direction. However, experimental data have clearly shown that resistivity along the $y$ direction is larger than the $x$ direction~\cite{Chu10Science, Ishida13PRL}. This counterintuitive behavior in the AFM phase of the 122 systems is naturally explained if one takes into account both anisotropic Fermi surfaces and nonmagnetic-impurity scattering~\cite{Sugimoto14PRB}.

In the 11 system, the ordering vector is close to $\mathbf{Q}=(\pi/2,\pi/2)$~\cite{Bao09PRL,Li09PRB,Lipscombe11PRL}, unlike the 122 systems. The ordering is called double stripes, where the AFM spin arrangement occurs along one of the second-neighbor Fe-Fe directions and the FM arrangement appears perpendicular to the AFM-ordered direction. We call the AFM (FM) direction the $a$ ($b$) direction [see Fig.~\ref{fig:spin-config}(b)]. A recent experiment has shown that resistivity along the $a$ direction (AFM direction) is larger than the $b$ direction (FM direction)~\cite{Liu15PRB}, which is opposite to the 122 systems and the intuitive view looks accurate. However, it should be examined carefully whether such an intuitive view is really accurate. In order to understand in-plane anisotropy systematically, the procedure applied to the 122 systems in the previous study~\cite{Sugimoto14PRB} would be helpful. 

In this paper, we theoretically examine in-plane resistivity in the AFM phase of the 11 system at zero temperature. The AFM state is obtained by a mean-field theory of a five-orbital Hubbard model. The anisotropy of resistivity is obtained by a recently developed multi-orbital memory function approach~\cite{Sugimoto14PRB} that takes into account nonmagnetic impurity scattering. In the approach, resistivity is proportional to scattering rate divided by Drude weight. Calculated results are consistent with experimental data, showing the resistivity in the AFM-ordered direction larger than that in the FM-ordered direction.
In contrast to the 122 systems~\cite{Sugimoto14PRB}, the anisotropy of resistivity is never reversed, though its magnitude may be changed as doping related to $x$, due to a transition of Fermi surface topology.
As a result of the contribution from Drude weight and scattering rate reflecting the electronic band structure at the Fermi level, the anisotropy remains opposite to the 122 systems.
Finding out that the anisotropy is attributed to Fermi surface topology of the ordered phase, we derive a conclusion that the intuitive view based on an arrangement of local spins is unlikely to be a good starting point as expected in metallic systems.

\section{Formulation}
We introduce a multiband Hubbard Hamiltonian for $d$-electron system in two-dimensional square lattices, $H_d = H_0 + H_I$, which describes iron pnictides.
The noninteracting Hamiltonian $H_0$ is given by
\begin{eqnarray}
\hspace{-0.7cm}H_0=\sum_{i,j}\sum_{\mu,\nu,\s}
\sbra{
 t(\mathbf{\Delta}_{ij};\mu,\nu)
+\varepsilon_{\mu}\delta_{\mu,\nu}}
c_{i\mu\s}^\dagger c_{j \nu\s},
\end{eqnarray}
where $c_{i\nu\s}^\dagger$ creates an electron of an orbital $\mu$ with a spin $\s$ at the $i$-th Fe site with on-site energy $\varepsilon_{\mu}$.
The hopping energy $t(\mathbf{\Delta}_{ij};\mu,\nu)$ is for the one from the orbital $\nu$ at the site position $\mathbf{r}_j$ to $\mu$ at $\mathbf{r}_i$ between the sites distanced by $\mathbf{\Delta}_{ij}\equiv \mathbf{r}_i-\mathbf{r}_j$.
The interaction Hamiltonian $H_I$ can be written {\cred as follows, by assuming that the pair hopping equals the Hund coupling $J$}~\cite{Oles83PRB}:
\begin{align}
H_I =& U \sum_{i,\mu}
n_{i \mu \up} n_{i \mu \dn}
+
\paren{U-2J}
\sum_{i,\mu\neq\nu}
n_{i \mu \up} n_{i \nu \dn}
\nonumber\\
&-J\sum_{i,\mu\neq\nu}
(c_{i \mu\up}^\dagger c_{i \mu\dn}
c_{i \nu\dn}^\dagger c_{i \nu\up}
-
c_{i \mu\up}^\dagger c_{i \nu\up}
c_{i \mu\dn}^\dagger c_{i \nu\dn})
\nonumber\\
&+
\frac{U-3J}{2}\sum_{i,\mu\neq\nu,\s}
n_{i \mu\s}n_{i \nu\s}
,
\end{align}
where $n_{i \mu \up}=c_{i \mu\up}^\dagger c_{i \mu\up}$ and $U$ is the intraorbital Coulomb interaction.

In practice, we calculate the electronic state within the mean-field approximation by self-consistently solving the mean-field equations containing the AFM order parameter.
The order parameter is defined by
$\<n_{l\mathbf{Q}\,\mu\nu\s}\>=N^{-1}\sum_{\mathbf{k}}
\<c_{\mathbf{k}+l\mathbf{Q}\,\mu \s}^\dagger c_{\mathbf{k}\,\nu \s}\>
$
with $\mathbf{Q}$ being the ordering vector, $N$ being the number of the lattice points, and the Fourier transform $c_{\mathbf{k} \mu\s} = N^{-\frac{1}{2}} \sum_{i} c_{i \mu\s} \exp{\paren{\i \mathbf{k}\cdot \mathbf{r}_i}}$.
The ordering vectors arising from the spin configuration (Fig.~\ref{fig:spin-config}) are $(\pi,0)$ for the 122 system and $(\pi/2,\pi/2)$ for the 11 system~\cite{Bao09PRL,Li09PRB,Lipscombe11PRL}: Thus, the first Brillouin zone is reduced into $1/N_Q$, where $N_Q=4$ for the 11 system.
The multiplier $l$ in $l\mathbf{Q}$ takes $0,1,.\,.\,.,N_Q-1$.
Finding the solution satisfying the self-consistent condition
$
\sum_{\mathbf{k}, \epsilon}\psi^*_{\mu\epsilon\s}(\mathbf{k})
\psi_{\nu\epsilon\s}(\mathbf{k}+l\mathbf{Q})
=N \<n_{l\mathbf{Q}\,\mu\nu\s}\>
$,
we finally obtain a quasiparticle state of a band $\epsilon$, 
$
\gamma^\dagger_{\mathbf{k}\epsilon\s} =\sum_{l}\sum_{\mu} \psi^*_{\mu\epsilon\s}(\mathbf{k}+l\mathbf{Q}) c^\dagger_{\mathbf{k}+l\mathbf{Q}\mu\s}
$
with energy $E_{\mathbf{k\epsilon\s}}$.
The mean-field Hamiltonian $H^{\mathrm{MF}}$ is expressed as
$H^{\mathrm{MF}}=\sum_{\mathbf{k}_0,\s}E_{\mathbf{k}_0\epsilon\s} \gamma^\dagger_{\mathbf{k}_0\epsilon\s} \gamma_{\mathbf{k}_0\epsilon\s}$,
where $\mathbf{k}_0$ is restricted within the reduced zone.

We refer to the data from an {\it ab initio} model based on the downfolding scheme~\cite{Miyake10JPSJ} for the on-site energies and hopping integrals of FeTe.
Since Fermi surface topology in the paramagnetic phase does not fit to a nesting condition for $\mathbf{Q}=(\pi/2,\pi/2)$, we need to use stronger Coulomb interactions as compared to those of the 122 systems in order to stabilize the $\mathbf{Q}=(\pi/2,\pi/2)$ magnetic order. In fact, we get the order with magnetic moment $m=2.67\mu_B$ ($\mu_B$ is the Bohr magneton) at electron density $n=6.0$ corresponding to $x=0$ by setting $U=1.6$~eV and $J=0.32$~eV, which are larger than the values for the 122 systems used before ($U=1.2$~eV, $J=0.22$~eV, and $m=0.8 \mu_B$). A tendency toward a large value of $U$ and $J$ for the 11 system is consistent with {\it ab initio} low-energy models based on a constrained random-phase approximation~\cite{Miyake10JPSJ}.  The obtained $m$ is close to an experimental value of $m=2.25~\mu_B$~\cite{Li09PRB} as well as a theoretical value of $m=2.5~\mu_B$~\cite{Ma09PRL} obtained by an {\it ab initio} calculation based on the local spin density approximation and a value of $m=2.1~\mu_B$~\cite{Yin11NATM} obtained by a combined density-functional and dynamical mean-field theory. Fermi surfaces in our calculation are qualitatively similar to the {\it ab initio} calculation~\cite{Ma09PRL} in the sense that there are two components in the magnetic Brillouin zone at $n=6$ [see Fig.~\ref{fig:vf}(a)].
Naturally assuming that excess iron of concentration $x$ introduces electrons in the Fe plane, we change $n$ from 6.0 to 6.2 and obtain the double-striped AFM order.

To investigate the anisotropy of the electronic transport, we evaluate the resistivity, Drude weight, and scattering rate in the each direction along and across the stripes, i.e., the $b$ and $a$ directions, respectively.
Recently a multiorbital memory function technique that is a multiorbital version of the memory function theory~\cite{Gotze72PRB} has been developed~\cite{Sugimoto14PRB}, where nonmagnetic impurity is a source of elastic scattering and a Born approximation is employed. Within the method, resistivity along the $\alpha$ direction is given by the ratio of the imaginary part of the memory function $M''_\alpha$ to the charge stiffness or Drude weight $D_\alpha$:
\begin{align}
\rho_\alpha
=&
\frac{\paren{\hbar/N_{\mathrm{F}}} \sum_{\mathbf{k}_\mathrm{F}}M''_{\alpha}(\mathbf{k}_\mathrm{F})}
{2 D_\alpha},
\label{rho}
\end{align}
where $\mathbf{k}_\mathrm{F}$ represents $\mathbf{k}$ points at the Fermi level $E_\mathrm{F}$ and $N_{\mathrm{F}}$ is the number of the points.

In the calculation of $\rho_\alpha$ of the multi-orbital system, the memory function approach is rather simple and feasible, while the application of the Boltzmann equation to the multiorbital systems is limited and still within a phenomenological level~\cite{Gotze72PRB}
This is why we here adopted the memory function approach.

{$D_\alpha$ and $M''_\alpha$ are calculated from the current matrix $\mathcal{J}_{\epsilon,\epsilon'}$ and the impurity matrix $\mathcal{I}_{\epsilon,\epsilon'}$, which arise from the current operator $\mathbf{j}=-c\paren{\frac{\partial H}{\partial \mathbf{A}}}_{\mathbf{A}=0}$ ($c$ is the velocity of light) and the impurity Hamiltonian $H_{\mathrm{imp}}=I_{\mathrm{imp}}\sum_{\mu,\s}c_{\ell \mu\s}^\dagger c_{\ell \mu\s}$, respectively ---we here assume a nonmagnetic local potential $I_{\mathrm{imp}}$ at a site $\ell$ and hereafter set $r_{\ell}=0$.

The current matrix is defined as
\begin{align}
\mathbf{j} =\sum_{\epsilon,\epsilon',\s}
\mathcal{J}_{\epsilon,\epsilon'} \gamma^\dagger_{\mathbf{k}_0\epsilon\s} \gamma_{\mathbf{k}_0\epsilon\s},
\end{align}
and the impurity matrix is defined as
\begin{align}
H_{\mathrm{imp}} =\frac{1}{N}\sum_{\epsilon,\epsilon'} \mathcal{I}_{\epsilon,\epsilon'}(\mathbf{k}_0,\mathbf{k}'_0) \gamma^\dagger_{\mathbf{k}_0\epsilon\s} \gamma_{\mathbf{k}'_0\epsilon\s}.
\end{align}
The $\alpha$ component of $\mathcal{J}_{\epsilon,\epsilon'}$ is calculated as
\begin{align}
\mathcal{J}_{\epsilon,\epsilon'}^{(\alpha)}(\mathbf{k}_0,\s)
&=
\frac{\i}{N}\frac{e}{\hbar}
\sum_{l,i,j,\mu,\nu}
\paren{\mathbf{a}_\alpha\cdot\mathbf{\Delta}_{ij}}
t(\mathbf{\Delta}_{ij};\mu,\nu)\nonumber\\
&\times
\exp\sbra{\i(k_0+l\mathbf{Q})\cdot \mathbf{\Delta}_{ij}}
\nonumber\\
&\times 
\psi^*_{\mu \epsilon \s}(\mathbf{k}_0+l\mathbf{Q})
\psi_{\nu \epsilon' \s}(\mathbf{k}_0+l\mathbf{Q}),
\end{align}
where $e$ is the elementary charge and $\mathbf{a}_\alpha$ is a unit vector pointing to the $\alpha$ direction.
The impurity matrix is calculated as
\begin{align}
\mathcal{I}_{\epsilon,\epsilon'}^{(\alpha)}(\mathbf{k}_0,\mathbf{k}'_0,\s)
&=
I_{\mathrm{imp}}\sum_{\mu,l,l'}
\psi^*_{\mu \epsilon \s}(\mathbf{k}_0+l\mathbf{Q})
\psi_{\mu \epsilon' \s}(\mathbf{k}'_0+l'\mathbf{Q}).
\end{align}

Drude weight is obtained from
\begin{align}
D_{\alpha}
&=
\frac{1}{N_{\mathrm{F}}}
\sum_{\epsilon_\mathrm{F},\s}\sum_{\mathbf{k}_\mathrm{F}}
\frac{1}{\abs{v(\mathbf{k}_\mathrm{F})}}
\abs{\mathcal{J}_{\epsilon_\mathrm{F},\epsilon_\mathrm{F}}^{(\alpha)}(\mathbf{k}_\mathrm{F},\s)}^2
\end{align}
where the set $(\epsilon_\mathrm{F},\mathbf{k}_\mathrm{F},\s$) is chosen so that $E_{\mathbf{k}_\mathrm{F}\epsilon_\mathrm{F}\s}=E_\mathrm{F}$ and $v(\mathbf{k}_\mathrm{F})$ is the Fermi velocity at $\mathbf{k}_\mathrm{F}$.
Scattering rate can be evaluated as
\begin{align}
M''_{\alpha}(\mathbf{k}_\mathrm{F})
=&
\frac{\pi n_c}{2D_\alpha}
\frac{1}{N'_{\mathrm{F}}}
\frac{1}{\abs{v(\mathbf{k}_\mathrm{F})}}\nonumber\\
&\times
\sum_{\epsilon_\mathrm{F},\epsilon'_\mathrm{F},\s}
\sum_{\mathbf{k}'_\mathrm{F}}
\frac{1}{\abs{v(\mathbf{k}'_\mathrm{F})}}
\abs{\mathcal{A}^{(\alpha)}_{\epsilon_\mathrm{F},\epsilon'_\mathrm{F}}(\mathbf{k}_\mathrm{F},\mathbf{k}'_\mathrm{F},\s)}^2,
\label{M}
\end{align}
where 
\begin{align}
\mathcal{A}^{(\alpha)}_{\epsilon,\epsilon'}(\mathbf{k},\mathbf{k}',\s)
=&
\mathcal{I}_{\epsilon,\epsilon'}^{(\alpha)}
(\mathbf{k},\mathbf{k}',\s)
\nonumber\\
&\times\sbra{
\mathcal{J}_{\epsilon,\epsilon}^{(\alpha)}
(\mathbf{k},\s)
-
\mathcal{J}_{\epsilon',\epsilon'}^{(\alpha)}
(\mathbf{k}',\s)}.
\label{A}
\end{align}

It can be seen how Fermi velocities affect scattering rate by taking into account $\mathcal{J}^{(\alpha)}(\mathbf{k}) \propto v_\alpha(\mathbf{k})$, where the subscript $\alpha$ means its $\alpha$ component.
From Eqs.~(\ref{M}) and (\ref{A}), $M''_\alpha(\mathbf{k}_\mathrm{F})$ is contributed from Fermi velocities as follows:
\begin{align}
M''_\alpha(\mathbf{k}_\mathrm{F})&\propto
\frac{1}{D_\alpha}\sum
\abs{\mathcal{I}_{\epsilon,\epsilon'}^{(\alpha)}(\mathbf{k},\mathbf{k}',\s)}^2
\frac{\abs{v_\alpha(\mathbf{k}_\mathrm{F})-v_\alpha(\mathbf{k}'_\mathrm{F})}^2}
{\abs{v(\mathbf{k}_\mathrm{F})}\,\abs{v(\mathbf{k}'_\mathrm{F})}}
\label{M''vs.vF}
\end{align}
Roughly speaking, $M''_\alpha$ can be enhanced for $\mathbf{k}_\mathrm{F}$ with $v(\mathbf{k}_\mathrm{F})$ in the $\alpha$ direction, and diminished for a large $v(\mathbf{k}_\mathrm{F})$.

\section{results}

\begin{figure}[t]
\begin{center}
\includegraphics[width = 1.0\linewidth]{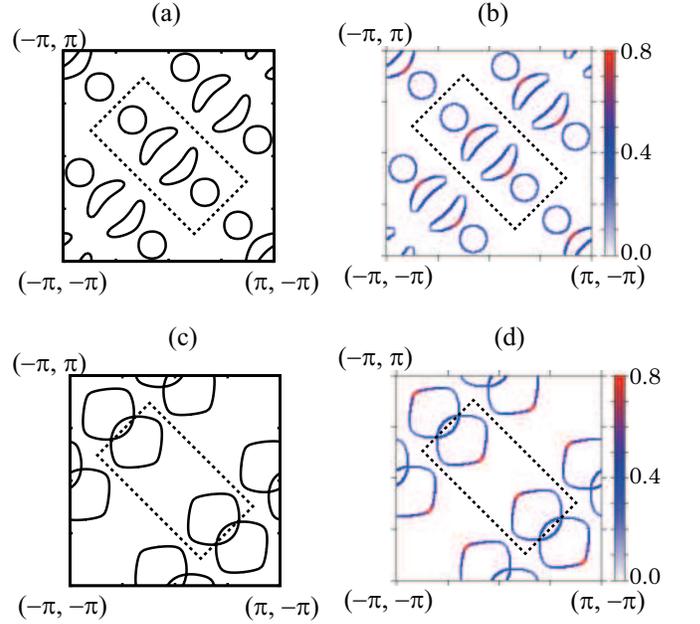}
\caption{(Color online) The Fermi surface and magnitude of Fermi velocity $\abs{v_{\mathrm{F}}}$.
Plotted are (a) the Fermi surface and (b) $\abs{v_{\mathrm{F}}}$ for $n=6.0$; panels (c) and (d) show the same for $n=6.2$.  The rectangles are the reduced zone.
}
\label{fig:vf}
\end{center}
\end{figure}

We now present the results of the calculation.
First, we examine Fermi velocity as a fundamental of the transport property together with features of the Fermi surface.
Figure~\ref{fig:vf} shows the distribution of the velocities $\abs{v_\mathrm{F}}$ [(b) and (d)] on the Fermi surface [(a) and (c)] for different electron densities: $n=6.0$ and $6.2$.
In the undoped case, the Fermi surface has crescent-shaped hole pockets and circular electron pockets.
The former has the largest $\abs{v_\mathrm{F}}$ on the side facing the $b$ direction.
As electrons are doped, the hole pockets shrink and vanish, while the electron pockets with their radii increased grow into an interlocking structure.
The $n=6.2$ case is shown in Fig.~\ref{fig:vf} (c).
In both cases, the largest velocity is directed to $\pm b$ rather than $\pm a$.
This affords a preference in conductive direction for $b$ over $a$.
The contribution of velocities to the transport is closely reflected in that of Drude weight through the current operator.
The preference for $b$ conduction is, therefore, interpreted directly as an effect of $D_\alpha$ in Eq.~(\ref{rho}):
The larger the velocity along $b$, the larger the value of $D_b$ and the smaller the resistivity $\rho_b$ along $b$.
We find, as a result, that the Fermi velocity feature tends to increase the $b$ conduction (or decrease the $a$ resistivity):
This is consistent with the experimental results.

\begin{figure}[t]
\begin{center}
\includegraphics[width = 1.0\linewidth]{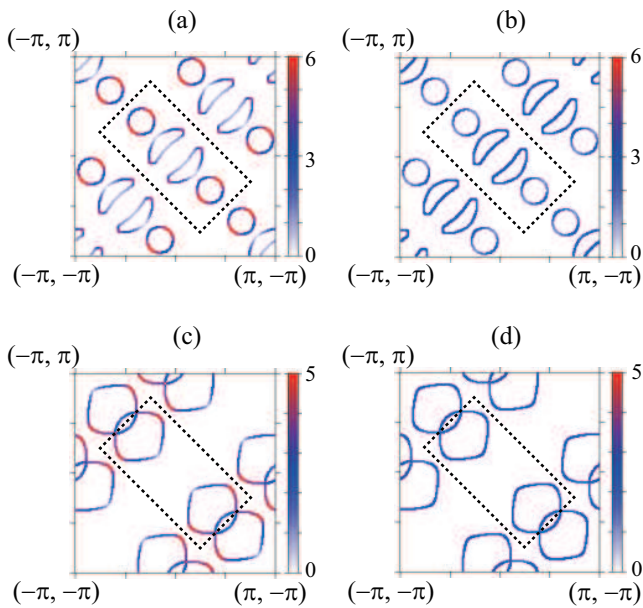}
\caption{(Color online) Scattering rate arising from the memory functions $M''_{a}$ (a, c) and $M''_{b}$ (b, d) for the cases of $n=6.0$ (a, b) and $n=6.2$ (c, d). 
The rectangles are the reduced zone.
}
\label{fig:memfn}
\end{center}
\end{figure}

Since we have perceived the effect of $D$ in Eq.~(\ref{rho}) on $\rho$, we next focus on that of $M''$, i.e., the scattering rate as an effect of impurities to the transport property.
The intensity of the scattering rate is represented by $M''$ at the Fermi level [Eq.~(\ref{M})].
Since the formula of $M''_\alpha$ has a factor of inverse $D_\alpha$, it is basically expected to behave in an opposite way: The ratio $M''_a/M''_b$ tends to increase as $D_a/D_b$ decreases ($D_b/D_a$ increases).

Another factor to determine $M''_\alpha$ is $\abs{v(\mathbf{k}_\mathrm{F})}$, which relates to its dependence on $\mathbf{k}_\mathrm{F}$.
As mentioned above [Eq.~(\ref{M''vs.vF})], $M''_b$ is enhanced for $\mathbf{k}_\mathrm{F}$ with $v(\mathbf{k}_\mathrm{F})$ in the $b$ direction, and diminished for a large $v(\mathbf{k}_\mathrm{F})$: The result shown in Fig.~\ref{fig:memfn} is consistent with this basic aspect.
In both cases $n=6.0$ and $6.2$, scatterings in the $a$ direction mostly coming from the circular pockets overwhelm those in $b$.

In total, it results in $M''_a > M''_b$ as shown in Fig.~\ref{fig:ratio}, where it is also demonstrated that the anisotropy of $D$ is larger than that of $M''$.
As a result, $M''$ contributes to the anisotropy $\rho_a>\rho_b$ as well as $D$ in a way that it is much less than that of $D$.

\begin{figure}[t]
\begin{center}
\includegraphics[width = 0.7\linewidth]{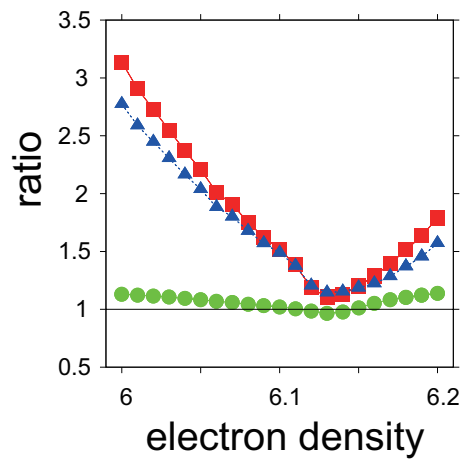}
\caption{(Color online) Anisotropy of the resistivity. Plotted are the ratio of $\rho_a/\rho_b$ (squares), $D_b/D_a$ (triangles), and $M''_a/M''_b$ (circles).
}
\label{fig:ratio}
\end{center}
\end{figure}

Hence, we obtain the ratio of $\rho_a/\rho_b$ consistent with the experimental result $\rho_a>\rho_b$.
In addition, this turns out to contribute to the anisotropy of $D$.
$D$ closely reflects the structure of Fermi pockets, and so does $\rho_a/\rho_b$.

The anisotropy $\ds{\rho_a>\rho_b}$ is never reversed despite doping. This is different from the 122 systems, where the anisotropy reverses in hole doping~\cite{Sugimoto14PRB}.
As to the doping effect in the 11 system, the fundamental properties are unchanged in terms of anisotropy unless the structure of Fermi pockets is changed.
This occurs around $n=6.13$, where the crescent-shaped electron pockets shrink to vanish, and the circular hole pockets come to link with each other.
As the hole pockets grow with doping, the ratio $\rho_a/\rho_b$ decreases until $n=6.13$ and increases for $n>6.13$, where the hole pockets are linked pairwise.
Hence, the anisotropy is never reversed by doping.

Experimental data indicate that $\rho_a/\rho_b$ is more than unity but roughly less than 1.3~\cite{Liu15PRB}. Such a range is roughly located around $n=6.13$ in Fig.~\ref{fig:ratio}.  This $n$ larger than $6.1$ is not unrealistic since excess Fe concentration is near $x=0.08$~\cite{Liu15PRB}, and thus $0.08m$ electrons will be added to $n=6$ assuming Fe$^{m+}$ for the excess Fe.

\section{Conclusion}
We have investigated the origin of the anisotropic resistivity based on calculations of memory function and Drude weight.
From the calculation, it is revealed that the anisotropic property mostly arises from that of Drude weight, which is closely related to Fermi velocity.
Since the anisotropy of Drude weight directly represents that of electronic states at the Fermi level, we simply understand that the anisotropic resistivity originates from the anisotropic Fermi surface caused by the magnetic order.
We have reached this simple interpretation without introducing any bold, hypothetical assumption.
This means that the symmetry breaking induced by the magnetic order directly appears in the transport property ---not through the spin configuration, but through the Fermi surface topology.
This is important in advancing the study of the 11 system.

\begin{acknowledgments}
This work was supported in part by MEXT as a social and scientific priority issue (Creation of new functional devices and high-performance materials to support next-generation industries) to be tackled by using post-K computer and by Grant-in-Aid for Scientific Research from the Japan Society for the Promotion of Science (Grants No. 26287079 and No. 26400381). 
\end{acknowledgments}

\nocite{*}



\end{document}